# Indirect coupling between localized magnetic moments in zero-dimensional graphene nanostructures (quantum dots)

K. Szałowski[*]

Department of Solid State Physics, Faculty of Physics and Applied Informatics, University of Łódź,
ulica Pomorska 149/153, PL 90-236 Łódź, Poland

The indirect Ruderman-Kittel-Kasuya-Yosida (RKKY) coupling between on-site magnetic impurities is studied for two kinds of graphene nanoflakes consisting of approximately 100 carbon atoms, posessing either zigzag or armchair edge. The tight-binding Hamiltonian with Hubbard term is used in non-perturbative calculations of coupling between the impurities placed at the edge of the structure. In general, for zigzag-edged nanoflakes a pronounced coupling robust against charge doping is found, while for armchair-edged structures the interaction is weaker and much more sensitive to charge doping. Also the distance dependence of indirect exchange differs significantly for both edge forms.

## 1. Introduction

Emergence of spin-based electronics requires search for appropriate building blocks for spintronic devices. One of the candidates may be zero-dimensional, few-atom magnetic nanostructures. Progress in their understanding and design requires extensive knowledge of the underlying magnetic interactions, which at present can be directly measured in experiment [1]. One of the promising platforms for development of such devices seems graphene. Its unique electronic properties result in highly nontrivial magnetic properties. In particular, zero-dimensional graphene nanostructures (nanoflakes, quantum dots) offer the possibility to design their electronic structure and thus adjust their spintronic functionalities [2-7]. Moreover, it is possible to tune their properties by charge doping, extending the possibilities offered by more traditional quantum dots [8].

All the mentioned factors strongly stimulate the studies of magnetism in graphene nanostructures. One of the related subjects is an indirect RKKY coupling between localized magnetic moments in geometrically confined graphene systems. The recent works deal with zero-dimensional [9,10] and one-dimensional [11] nanostructures, in addition to the extensive studies of RKKY interaction in infinite graphene layers ([12-15]).

In our paper we present the numerical calculations of RKKY coupling between on-site localized spins in two types of graphene nanoflakes (GNFs), concentrating on the importance of the edge form. To supplement our earlier considerations [9,10], we study larger structures, each consisting of approximately 100 carbon atoms.

## 2. Theoretical model

In our work we focus the attention on two forms of GNFs (see insets in Fig. 1). One of them is characterized by zigzag edge, while the other has armchair edge. Both nanoflakes are in general of triangular shape and their size is described by the number $M$ of hexagons forming their edge.

The electronic structure of the GNFs is described by means of a nearest-neighbour tight-binding Hamiltonian supplemented with a mean-field Hubbard term [2,4,9,10,13,16]. The presence of a pair of on-site impurity spins is assumed. The spins of these magnetic impurities at sites $a$ and $b$ (belonging to two interpenetrating sublattices) interact with charge carrier spins with exchange energy $J$. In the present paper the normalized values of $J/t = 0.1$ and $U/t = 1.0$ were assumed.

$$H = -t \sum_{\langle i,j \rangle, \sigma=\uparrow,\downarrow} (c^\dagger_{i,\sigma} c_{j,\sigma} + c^\dagger_{j,\sigma} c_{i,\sigma}) + U \sum_{i,\sigma} n_{i,\sigma} \langle n_{i,-\sigma} \rangle$$
$$- U \sum_i \langle n_{i,\uparrow} \rangle \langle n_{i,\downarrow} \rangle + J(S_a s^z_a + S_b s^z_b)$$

The Hamiltonian is diagonalized self-consistently in single-particle approximation in presence of a constant number of charge carriers [11-15]. The charge doping is described by the filling level of discrete energy states, where half-filling corresponds to the charge neutrality in graphene. The magnitude of an indirect RKKY coupling between impurity spins is proportional to the difference in total energy of the system for antiparallel and parallel orientation of impurity spins $2S^2 J^{RKKY} = E^{AF}_{total} - E^F_{total}$ [9-13].
Let us emphasize that our approach is a non-perturbative one.

## 3. Numerical results

The dependence of the indirect coupling on the energy levels filling factor is presented in Fig. 1 for selected impurity positions (3rd and 4th nearest neighbours) along the edge of GNFs with $M = 9$. For a zigzag-edged GNF, the coupling in the vicinity of charge neutrality is strongly enhanced owing to the first-order perturbative mechanism involving spin-polarized states occupied by single electrons [9-11]. For impurities in the same sublattice the interaction is ferromagnetic, while it is antiferromagnetic for different sublattices [12,14,15]. This rule is broken only for considerable charge doping for GNFs of such size as those considered here.

[*]e-mail: kszalowski@uni.lodz.pl

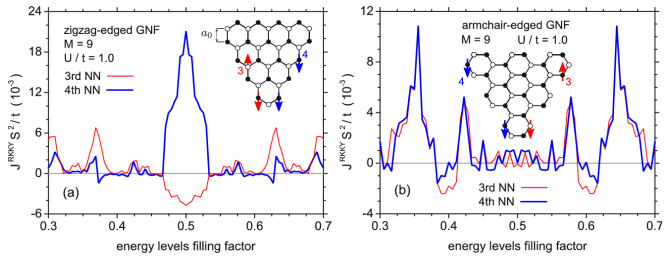

Fig.1. Dependence of an indirect coupling on energy levels filling factor for zigzag-edged GNF (a) and armchair-edged GNF (b), for 3rd and 4th NN edge impurities (see insets).

For an armchair-edged GNF, the interaction is relatively weak for charge neutrality conditions, however, its value and sign shows pronounced sensitivity even to small charge doping, even though the GNF is relatively large. For higher doping levels, pronounced peaks in coupling energy occur and the coupling behaviour becomes rather similar both for 3rd and 4th neighbour impurities, what indicates a strong deviation from the rule relating the coupling sign just with the spin placement in the same/different sublattices.

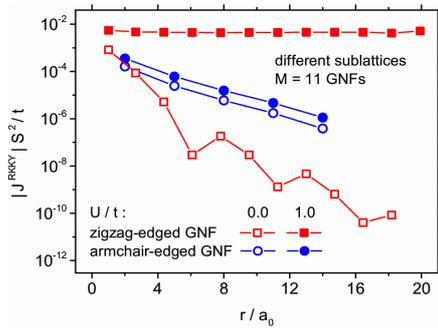

Fig.2. Distance depen-dence of an indirect coupling magnitude along the edge of the undoped GNFs with $M = 11$, for magnetic impurities in different sublattices.

The distance dependence of an indirect coupling magnitude for both considered GNFs (with $M = 11$) is plotted in Fig. 2 for impurities in different sublattices (so that the interaction is antiferromagnetic), positioned along the GNF edge. The case of undoped, charge-neutral structures is considered. It is visible that for an armchair-edged GNF, the coupling tends to decay exponentially with the distance along the edge and the influence of the Coulombic term is rather weakly pronounced. On the contrary, the interaction behaviour for a zigzag-edged GNF is much more sensitive to the presence of the Hubbard term. In its absence the interaction decays considerably faster and shows some magnitude oscillations. The on-site Coulombic interactions cause the coupling to become distance-independent along the edge of a triangular, zigzag-edged GNF, what can be attributed to the first-order indirect coupling mechanism mediated by zero energy states, which exhibit spontaneous magnetic polarization [2,4,10,13].

## 4. Final remarks

The presented calculations for considerably large nanoflakes confirm the importance of the edge form for the indirect coupling behaviour. The influence of charge doping was studied for a wide range of energy level filling factors. The relatively large size of the considered structures allowed also to investigate the distance dependence of coupling, yielding the distance-independent coupling at the zigzag GNF edge (induced by Hubbard term) and exponentially decaying coupling for an armchair one. This results show some directions for design of nanostructures with desired magnetic properties (robust or doping-sensitive coupling).

**Acknowledgements**

This work has been supported by Polish Ministry of Science and Higher Education on a special purpose grant to fund the research and development activities and tasks associated with them, serving the development of young scientists and doctoral students.

The computational support on HUGO cluster at P. J. Šafárik University in Košice is gratefully acknowledged.